\documentclass[aps,prb,showpacs,floatfix,twocolumn,superscriptaddress]{revtex4-2}
\usepackage{graphicx}
\usepackage{dcolumn}
\usepackage{cancel}
\usepackage{bm}
\usepackage{tikz}
\usepackage{siunitx}
\usepackage{tablefootnote}
\usepackage{physics}
\usepackage[section]{placeins}
\definecolor{darkblue}{rgb}{0,0,.65}
\definecolor{darkgreen}{rgb}{1,0,0}
\usepackage{nicefrac}
\usepackage{booktabs}
\usepackage[%
  pdfauthor={Benedikt Placke},%
  pdfstartview=FitH,%
  breaklinks=true,%
  bookmarks=true,%
  colorlinks=true,%
  anchorcolor=black,%
  citecolor=red,
  filecolor=black,%
  menucolor=black,%
  urlcolor=blue,%
  linkcolor=red,%
 ]{hyperref}

\usepackage[english]{babel}
\usepackage{bm}
\usepackage{amssymb}
\usepackage{color, soul}
\usepackage{amsmath}
\usepackage{amstext}
\usepackage{latexsym}
\usepackage{graphicx}
\usepackage{mathtools}
\usepackage[ruled]{algorithm2e}
\usepackage{enumitem}

\newcommand{\scrZ}{\mathcal{Z}}
\newcommand{\scrG}{\mathcal{G}}

\newcommand{\scrS}{\mathcal {S}}
\newcommand{\scrC}{\mathcal {C}}
\newcommand{\scrN}{\mathcal {N}}
\newcommand{\scrO}{\mathcal {O}}

\begin{document}
\title{ Percolation in the two-dimensional Ising model}

\author{Tao Chen}%
\thanks{These two authors contributed equally to this paper.}
\affiliation{
Hefei National Laboratory for Physical Sciences at the Microscale and Department of Modern Physics, University of Science and Technology of China, Hefei 230026, China}
\affiliation{Hefei National Laboratory, University of Science and Technology of China, Hefei 230088, China}

\author{Jinhong Zhu}%
\thanks{These two authors contributed equally to this paper.}
\affiliation{
Hefei National Laboratory for Physical Sciences at the Microscale and Department of Modern Physics, University of Science and Technology of China, Hefei 230026, China}

\author{Wei Zhong}
\thanks{zhongwei2284@hotmail.com}
\affiliation{Minjiang Collaborative Center for Theoretical Physics, College of Physics and Electronic Information Engineering, Minjiang University, Fuzhou, 350108 Fujian Province, PR China}

\author{Sheng Fang}%
\thanks{shengfang9503@bnu.edu.cn}
\affiliation{School of Systems Science, Beijing Normal University, 100875 Beijing, China}

\author{Youjin Deng}%
\thanks{yjdeng@ustc.edu.cn}
\affiliation{
Hefei National Laboratory for Physical Sciences at the Microscale and Department of Modern Physics, University of Science and Technology of China, Hefei 230026, China}
\affiliation{Hefei National Laboratory, University of Science and Technology of China, Hefei 230088, China}

\date{\today}

\begin{abstract} 
The study of the Ising model from a percolation perspective has played a significant role in the modern theory of critical phenomena. We consider the celebrated square-lattice Ising model and construct percolation clusters by placing bonds, with probability $p$, between any pair of parallel spins within an extended range beyond nearest neighbors. At the Ising criticality, we observe two percolation transitions as $p$ increases: starting from a disordered phase with only small clusters, 
the percolation system enters into a stable critical phase that persists over a wide range $p_{c_1} < p < p_{c_2}$, and then develops a long-ranged percolation order with giant clusters for both up and down spins. At $p_{c1}$ and for the stable critical phase, the critical behaviors agree well with those for the Fortuin-Kasteleyn random clusters and the spin domains of the Ising model, respectively. At $p_{c2}$, the fractal dimension of clusters and the scaling exponent along $p$ direction are estimated as $y_{h2} = 1.958\,0(6)$ and $y_{p2} = 0.552(9)$, of which the exact values remain unknown. These findings reveal interesting geometric properties of the two-dimensional Ising model that has been studied for more than 100 years. 

\end{abstract}

\maketitle
\section{Introduction}

Phase transitions and critical phenomena constitute one of the most fundamental areas of study in statistical physics, with broad applications spanning nearly every branch of physics~\cite{Stanley1971Phase}. Phase transition describes abrupt changes in a system's macroscopic state under variations of external parameters—such as temperature, pressure, or magnetic field, etc. For continuous phase transitions, physical quantities like specific heat and correlation lengths diverge near the critical point, exhibiting power-law scaling behavior known as criticality. The associated scaling exponents, termed critical exponents, uniquely characterize these singularities. 

A cornerstone of the modern theory of phase transitions and critical phenomena theory is the renormalization group (RG) theory~\cite{wilson1971renormalization,wilson1971renormalizationa,wilson1972critical,wilson1975renormalization}. It systematically classifies continuous transitions into universality classes. Systems within the same universality class have the same set of critical exponents, corresponding to a common fixed point 
in the parameter space, despite their differing microscopic details. For instance, gas-liquid transitions and order-disorder transformations in binary alloys share the same universality class with the Ising model. Universality is governed by overarching factors such as spatial dimensionality, interaction symmetry, and the range of interactions~\cite{Stanley1999}.

 In the study of critical phenomena, the Ising model, which Lenz and Ising originally proposed to explain the ferromagnetic-to-paramagnetic phase transition observed at the Curie temperature~\cite{RevModPhys.39.883}, stands as one of the most paradigmatic systems. The zero-field partition function of the model is defined as:
        \begin{equation}
            \scrZ  =  \sum_{\{s_i\}} e^{K\sum_{\langle ij \rangle} s_i s_j}. 
            \label{eq:Ising}
        \end{equation}
Here, $\sum_{\{s_i \} }$ and $\sum_{\langle ij \rangle}$ denote the summations running over all possible spin configurations and pairs of interactions. The parameter $K$ corresponds to the coupling strength, and each spin $s_i$ takes values $\pm 1$, representing ``up" or ``down" orientations, respectively. 

In 1944, Onsager used the transfer matrix method to obtain the exact solution of the partition function for the square-lattice Ising model under the zero magnetic field, which gives the critical threshold $K_c = \ln(1+ \sqrt{2}) /2$ and a critical exponent $\alpha = 0$ ~\cite{Onsager1944}. Thereafter, Yang resolved Onsager's conjecture on spontaneous magnetization, 
deriving the critical exponent $\beta=1/8$ under a weak external magnetic field approximation~\cite{CNYang1952}. These results cemented the Ising model's universality within the RG framework, with thermal and magnetic scaling exponents $y_t = 1 $ and $y_h = 15/8$, respectively. Subsequently, in the 1980s, advancements in Coulomb gas theory and conformal field theory enabled the prediction of nearly all critical exponents for the two-dimensional (2D) Ising model, profoundly deepening its theoretical understanding~\cite{Friedan1984, Drouffe1989, Blote1992}.

The Ising model has also been extensively studied from the geometric perspective. In 1936, Peierls analyzed domain walls—boundaries separating spin clusters of opposing orientations—in the 2D Ising model, demonstrating the existence of long-range order at low temperature~\cite{Peierls_1936}. Later, Fisher introduced the droplet model to describe scaling behavior near gas-liquid critical points using an ensemble of non-interacting droplets, which was analogized to spin clusters~\cite{Fisher1967}. In two dimensions, these spin clusters captured critical singularities such as the divergence of the correlation length, while they could not give a quantitative description of the power-law divergence of the susceptibility. Coniglio and collaborators later attributed this discrepancy to the geometric properties of spin clusters: their excessive density required the introduction of a bond probability  $p = 1- e^{-2K}$ between aligned neighboring spins to construct clusters~\cite{Coniglio1980clusters}. This adjustment successfully explained the divergence of the susceptibility.

More precisely speaking, the Ising model can be mapped to the geometric random-cluster model through the Fortuin-Kasteleyn (FK) transformation with the partition function 
\begin{equation}
\scrZ_{\rm FK} = \sum_{\scrG} p^{b(\scrG)}(1-p)^{E-b(\scrG)} q^{k(\scrG)} \; , 
\label{eq:FKIsing}
\end{equation}
with $q=2$.
Here, $\scrG$ denotes the set of all possible subgraphs, $b(\scrG)$ represents the total number of bonds, and $k(\scrG)$ is the number of clusters~\cite{fortuin1969,fortuin1972}. The parameter $E$ corresponds to the total number of edges. The occupied bond probability $p$ is related to the coupling strength $K$ by $p = 1-e^{-2K}$. The parameter $q=2$ in Eq.~\eqref{eq:FKIsing} arises from the $Z_2$ symmetry of the Ising model. This framework generalizes to arbitrary real numbers $q > 0$, defining the celebrated random-cluster model~\cite{grimmett2006random}. When $q\to 1$, it reduces to the bond percolation model. Considering a spin representation, bonds are placed with probability $p$ on edges connecting aligned spins, forming spin clusters known as FK clusters, which correspond to the geometric clusters mentioned in Ref.~\cite{Coniglio1980clusters}. The backward-forward transformation between spin configurations and FK representations has facilitated the development of efficient cluster Monte Carlo algorithms, such as the Swendsen-Wang (SW) algorithm. 
   
Outside their utility in designing efficient Monte Carlo algorithms, geometric representations of the Ising model offer several additional advantages. Compared to spin representations, geometric frameworks provide access to a broader spectrum of geometric observables—many of which lack direct analogs in spin-based descriptions. Examples include wrapping probabilities $R_n$~\cite{Pinson1994Critical,Arguin2002}, defined as the probability that at least one cluster percolates by wrapping around periodic boundaries in $n$ spatial directions. Geometric representations also enable deeper conceptual insights and exhibit richer critical phenomena. For instance, studies have shown that the FK-Ising model simultaneously hosts two upper critical dimensions $(d_c, d_p)=(4,6)$, while the upper critical dimension of the spin Ising model is $d_c =4$~\cite{Fang_2022,PhysRevE.107.044103}. Leveraging geometric methods, Duminil-Copin and collaborators rigorously established the continuity and sharpness of the Ising phase transition in three dimensions~\cite{AizenmanDuminilCopinVladas2015} and proved the triviality of critical behavior in $d =4$~\cite{Aizeman2021Marginal}. In two dimensions, geometric representations further serve as a foundational framework for conformal field theory~\cite{francesco2012conformal} and stochastic Loewner evolution~\cite{lawler2001introduction}, yielding exact solutions for critical exponents and scaling limits. 

Another well-known geometric representation of the Ising model 
is the so-called loop representation, which can be obtained 
from the high-temperature expansion technique. 
In 2D, these loops also correspond to the boundaries between the up and down spins of the Ising model on the dual lattice. 
Further, starting from the FK random clusters of the Ising model, a random loop configuration can be obtained via the loop-cluster coupling and 
the loop-cluster algorithm~\cite{zhang2020loop}.

\begin{figure}[t]
\centering
\includegraphics[width=1.0\linewidth]{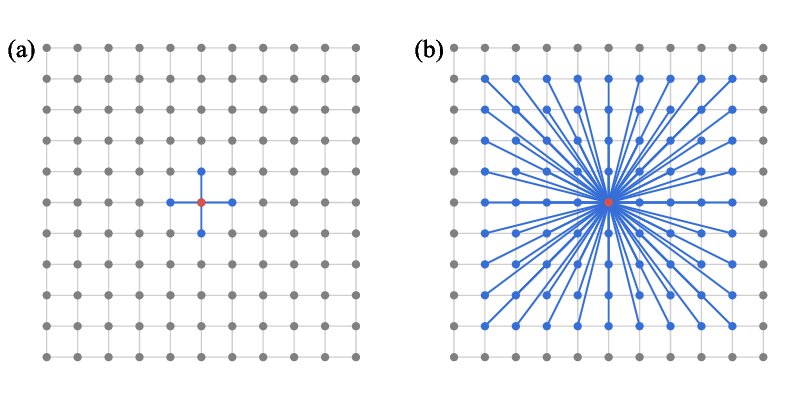}
\caption{Illustration of the equivalent-neighbor model for a given lattice site (red dot) and its neighbors (blue dot) with coordination number (a) $z_i=4$ and (b) $z_p=80$, respectively.}
\label{fig:coordination_number}
\end{figure}

In this work, we investigate percolation on the spin configuration of the celebrated square-lattice Ising model with two ingredients of generalization. 
First, as already in many previous studies~\cite{qian2005,Fang2021Percolation}, we lift the restriction 
of $p = 1 - e^{-2K}$ in the FK random cluster representation and 
treat $p$ as an independent parameter. 
Second, the range for placing bonds between parallel spins
is no longer restricted to nearest neighbors 
but extended to a large finite range.
A strategy of the so-called equivalent neighbor model is adopted~\cite{Ouyang2018}:
given a pair of lattice sites $(x_i,y_i)$ and $(x_j,y_j)$ 
of parallel spins, an occupied bond is placed with probability $p$ 
as long as $|x_i-x_j|\leq r$ and $|y_j-y_j|\leq r$ with $r>0$ 
being an integer. 
Figure~\ref{fig:coordination_number} illustrates the case in the current work, 
which has coordination numbers $z_i=4$ for the Ising interaction $(r_i=1)$ and $z_p =80$ for percolation ($r_p=4$).
In constructing percolation clusters from spin configurations, we segregate these configurations into ``majority" and ``minority" spin components. The ``majority" component corresponds to the part with the spin values that dominate the spin configuration, while the ``minority" component consists of the remaining spins. Percolation clusters are then formed within these distinct components. 

\begin{figure}[t]
    \centering
    \includegraphics[width=0.9\linewidth]{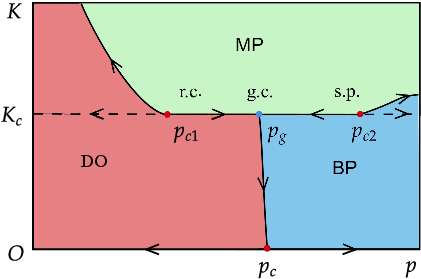}
    \caption{Phase diagram and renormalization group flows for percolation of the 2D Ising model in the $(p, K)$ parameter space, where $p$ is independent of $K$. Arrows indicate the renormalization group flows. Along the thermodynamic critical line $K=K_c$, the points labeled ``r.c." (random-cluster), ``g.c." (geometric-cluster), and ``s.p." (second percolation) correspond to the respective fixed points. The stable fixed point for the g.c. is marked as a blue dot, and the unstable fixed points are denoted by red dots. Solid lines represent phase boundaries. Different phases are indicated by colors: the disordered phase (DO), the phase with the majority spin configuration being percolated (MP), and the phase where both majority and minority spin configurations are percolated. (BP).
    }
\label{fig:phase_diagram}
\end{figure}

Extensive Monte Carlo simulations are performed and the obtained phase diagram in the $(K,p)$ plane is illustrated in Fig.~\ref{fig:phase_diagram}, where solid curves demarcate phase boundaries separating three distinct regimes: a disordered phase (DO), a majority-spin percolation phase (MP), and a bistable percolation phase (BP) where both majority and minority clusters percolate. Arrows denote RG flow trajectories, highlighting universality and scaling behavior. For small $p$ and $K<K_c$, the system resides in the disordered phase, characterized by Gaussian spin fluctuations. As the probability $p$ increases, both majority and minority spin clusters percolate at the same threshold, transitioning the system into the BP phase. In the long-range ordered phase ($K > K_c$), the majority spins predominate, their number greatly exceeding that of the minority spins, resulting in a significantly higher percolation threshold for minority clusters than majority clusters. All observed transitions within these two regimes adhere to the standard two-dimensional percolation universality class. However, along the critical line $K=K_c$, critical spin fluctuations drive the universality classes of the phase transitions away from the 2D percolation universality.

Utilizing the Coulomb gas theory, we analyze the phase transition from the DO phase to a critical state at the threshold $p_{c1}$, denoted as a random cluster (r.c.) fixed point. When the percolation interaction range matches the Ising spin interaction range, this critical state belongs to the 2D FK Ising universality class. The RG exponent along the $p$-direction corresponds to the red-bond exponent $y_{p1} = y_r = 13/24$, whose exact value can be derived from Coulomb gas theory or conformal field theory~\cite{Nienhuis1982, Friedan1984}. As the bond probability $p$ further increases, the system approaches a geometric cluster (g.c.) fixed point at $p_g$, characterized by exponents $y_{p, \rm{gc}} = -5/8$ and $y_{h, \rm{gc}} = 187/96$. Beyond this, an additional unstable fixed point (s.p.) emerges at the threshold $p_{c2}$, which governs the transition from the critical state to the ordered phase. 

\par 

Although the obtained phase diagram in Fig.~\ref{fig:phase_diagram} is rather simple and the revealed critical behaviors (except at the second percolation threshold $p_{c2}$) are pretty clear, some subtleties and challenges are worthy of discussion. 
First, unlike in previous studies with $z_i=z_p=4$~\cite{qian2005}, the exact mapping between the spin and the percolation configuration is now 
lacking in the whole $K-p$ plane, and no insight can be taken from the FK transformation. 
Particularly, 
since the percolation range $z_p=80$ is much larger than 
the Ising interaction range $z_i=4$ and the percolation system is far beyond being planar, it is a priori unknown whether the system would develop immediately after the first percolation threshold $p_{c1}$, 
a long-range order parameter with giant percolation clusters
and whether a novel universality would arise at $p_{c1}$. 
In other words, there is no guarantee that  
the second percolation threshold $p_{c2}$ and 
the critical phase for a wide range $p_{c1} < p < p_{c2}$ would exist. 
Second, the fractal dimensions for the FK random clusters and the spin domains are 
respectively $d_f({\rm r.c.})=15/8$ and $d_f({\rm g.c.})=187/96$ 
and their values are very close to the spatial dimensionality $d=2$. For finite systems, this means that the largest percolation cluster can occupy a large fraction of the lattice, 
making it difficult to distinguish a large but fractal cluster from a giant one. 
In Ref.~\cite{qian2005}, for the equivalent-neighbor Ising model with $z_i=z_p=20$, severe finite-size corrections are observed in conventional quantities 
that are associated with the distribution of cluster sizes.

To explore the phase diagram with reduced effects from finite-size corrections, we make use of the wrapping probabilities of percolation clusters, 
which can be regarded as topological observables. 
As the well-known Binder cumulant, these dimensionless probabilities 
provide a powerful tool to locate the phase transition points in finite-size scaling analysis. 
Further, by making proper combinations of different wrapping probabilities,  
the so-called critical polynomials can be defined, 
for which corrections-to-scaling can be dramatically reduced or even eliminated in many cases~\cite{Xu2021,Bao-Zong2021,Scullard2022,Scullard_2021,Lv_2018}. To locate the first percolation threshold $p_{c1}$, 
we use the critical polynomial $P_b = R_2 - 2 R_0$, 
where $R_2$ is the probability that there exists a percolation cluster to 
wrap around the periodic boundaries both along the $x$ and $y$ directions 
and $R_0$ is the probability that no wrapping cluster appears in either of the directions. 
It is known that, for the critical square-lattice Ising model ($z_p=z_i=4$), 
the exact value of $P_b$ at criticality is $P_{b,c}=0$ and finite-size corrections are completely absent.
Thus, we expect that the finite-size scaling of $P_b$ can serve as a powerful quantity 
to locate $p_{c1}$ and examine its universality. 
For the critical stable phase over the range $p_{c1} < p < p_{c2}$, we define the 
critical polynomial as $P_{bh} = (R_{\rm 2M} +R_{\rm 2m}-R_{\rm 0M} - R_{\rm 0m})/2$, 
where the wrapping probabilities of the percolation clusters 
are now measured separately for the major (M) and the minor (m) spins. 
For the spin domains ($p=1$) of the critical Ising model on the triangular lattice ($z_i=z_p=6$), 
the self-matching argument predicts the exact value $P_{bh,c}=0$. 
From the assumption of universality, one expects that $P_{bh,c}=0$  would hold true in
the critical stable phase for $p_{c1} < p < p_{c2}$ if it exists. 
Finally, as discussed above, it is very challenging to locate the second percolation threshold $p_{c2}$, since the fractal dimension $d_f$ is already close to $2$ in the critical stable phase.
After a series of numerical tries, we find that the critical polynomial, 
$P_{bh} = (R_{\rm 2M} +R_{\rm 2m}-R_{\rm 0M} - R_{\rm 0m})/2$, suffers from less 
finite-size corrections and its critical value at $p_{c2}$ is close to but not identical to 0.
 
Using finite-size scaling analysis, we determine the critical thresholds $p_{c1}$ and $p_{c2}$, the stable fixed point $p_g$, and their corresponding RG exponents. At these thresholds, we measure the fractal dimensions of the largest clusters, which coincide with the magnetic RG exponent $y_h$. For random cluster fixed point at $p_{c1}=0.020~325~7(5)$, we find $y_{p1}= 0.543(3)$ and $y_{h1} = 1.874~9(8)$, consistent with the red-bond exponent $y_r = 13/24$ and the magnetic exponent $y_h = 15/8$, respectively. The value of the critical polynomial agrees with the universal value $P_{b,c}=0$
for the 2D Ising universality class. 
For the geometric cluster fixed point at $p_g=0.035~3(6)$, we obtain $y_{p, \text{gc}}=-0.624(1)$ and $y_{h, \text{gc}}=1.948(3)$, aligning with predictions for geometric cluster criticality,
and the critical polynomial is well consistent with $P_{bh,c}=0$. These results demonstrate the robustness of the universality class even when the percolation coordination number $z_p$ differs from the Ising interaction range $z_i$. Finally, for second percolation fixed point at $p_{c2}=0.345~82(2)$, we estimate $y_{p2} = 0.552(9)$ and $y_{h2} = 1.958\,0(6)$, though the exact theoretical values for these exponents remain unresolved and need further study.

The remainder of this paper is organized as follows. 
In Sec.~\ref{sec:Algorithms}, we present the simulation details, and numerical results are given in Sec.~\ref{sec:results}. Finally, we give a discussion in Sec.~\ref{Sec:discussion}.

\section{Simulation and Sampled Quantities}
\label{sec:Algorithms}
We simulate the square-lattice Ising model ( $z_i=4$) at its critical coupling $K_c= \text{ln}(1+\sqrt{2})/2$ using the Swendsen-Wang (SW) algorithm~\cite{Swendsen1987} under periodic boundary conditions. For each spin configuration, bonds are probabilistically placed ($p$) between parallel spins using an extended coordination number $z_p=80$, constructing percolation clusters. Due to the large $z_p$, the first percolation threshold $p_{c1}$ becomes small, leading to inefficiencies when placing bonds sequentially. To optimize this, we implement an accelerated bond-placement method~\cite{Blote2002,Deng2005} that skips unoccupied bonds and directly targets the next candidate edge. The distance $i$ between adjacent occupied bonds follows
\begin{equation}      
    i=1+\left\lfloor\text{ln}(r)/\text{ln}(1-p)\right\rfloor,\nonumber
\end{equation}
where $r\in(0,1]$ is a uniformly distributed random number and $\left\lfloor\cdot \right\rfloor$ is the floor function. We repeat this process until all edges in this system have been decided. 
Through this technique, the number of visited edges decreases from $z_pV/2$ to $pz_pV/2$ approximately, significantly improving the efficiency for small $p$.   

We performed extensive Monte Carlo sampling for system sizes ranging from $L=8$ to 512. For each system size $L$, measurements were taken across a series of bond occupation probabilities $p$. To ensure adequate statistics, we collected at least $10^7$ samples for each $L \leq 128$ and $10^6$ samples for $128 < L \leq 512$. Near the critical points $p_{c1}$ and $p_{c2}$, the number of samples was increased to more than $10^8$ for all system sizes up to $L=512$ to achieve high statistical precision. When sampling the quantities, lattice sites are partitioned into two subsets based on spin alignment: i.e., the ``majority" spins, which are the dominant orientation occupying most lattice sites, and the ``minority" spins are the less prevalent orientation. Based on this division, we sampled the following observables:

\begin{itemize} 
    \item The wrapping probabilities $\mathcal{R}_0$ and $\mathcal{R}_2$: For a given configuration and considering all clusters, if at least one cluster wraps around the lattice in both $x$ and $y$ directions, we set $\mathcal{R}_2 = 1$; otherwise, $\mathcal{R}_2 = 0$. If no cluster wraps around the lattice in any direction, we set $\mathcal{R}_0 = 1$; otherwise, $\mathcal{R}_0 = 0$. We also define the indicators $\mathcal{R}_{\text{0M}}$ and $\mathcal{R}_{\text{2M}}$ for majority clusters, and $\mathcal{R}_{\text{0m}}$ and $\mathcal{R}_{\text{2m}}$ for minority clusters, with the subscript ``M" denoting majority and ``m" denoting minority.
    \item The size of the largest cluster $\scrC_1$; 
    \item The  number of clusters $\scrN(s)$ with size $[s, s + \Delta s)$; 
    \item The second and fourth moment of all clusters $\scrS_2 = \sum_i \scrC_i^2$ and $\scrS_4 = \sum_i \scrC_i^4$, where $\mathcal{C}_i$ denotes the $i$th largest cluster.
\end{itemize}
From these observables, we calculate the following quantities
\begin{enumerate}[label=(\roman*)]
      \item  The critical polynomials

    \begin{equation}
    \begin{aligned}
        P_{b}&=\langle \mathcal{R}_{\text{2}}-2\mathcal{R}_{\text{0}}\rangle,\\
        P_{bh}&=\frac{1}{2}\langle \mathcal{R}_{\text{2M}}+\mathcal{R}_{\text{2m}}-\mathcal{R}_{\text{0M}}-\mathcal{R}_{\text{0m}}\rangle,\\
        P_{b2}&=\langle 2\mathcal{R}_{\text{2m}}-\mathcal{R}_{\text{0m}}\rangle,
    \end{aligned}
    \end{equation}
    where $\langle\cdot\rangle$ denotes the ensample average.
    \item The average value of the largest cluster $C_1 = \langle \scrC_1 \rangle$; 
    \item The cluster number density $n(s, L) = \frac{1}{L^d \Delta S} \langle \scrN(s) \rangle$; 
    \item  The Binder ratio
    \begin{equation}
        Q_{s} = \frac{\langle \scrS_2 \rangle^2}{ \langle 3 \scrS_2 ^2 - 2 \scrS_4  \rangle }. 
    \end{equation}    
\end{enumerate}

\section{RESULTS}
\label{sec:results}

\begin{figure*}[htpb]
    \centering
    \includegraphics[width=1.0\linewidth]{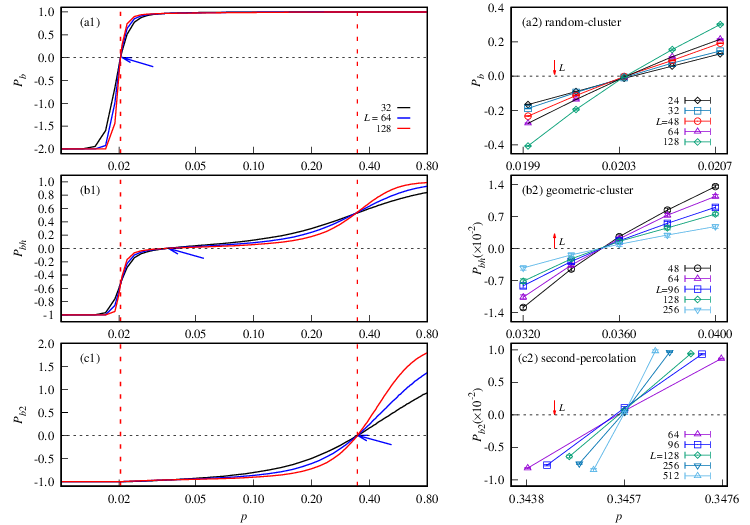}    
    \caption{Plots of critical polynomials $P_b, P_{bh}$ and $P_{b2}$ versus bond probability $p$ for equivalent-neighbor percolation with coordination number $z_p=80$ on critical Ising configurations at $K_c = \ln(1+\sqrt{2})/2$ for various system sizes $L$. Subfigures (a1)-(c1) display their scaling behavior with $p$ ranging from 0 to 0.8. 
    The blue arrows in each subfigure (a1), (b1) and (c1) denote the approximate location of the random-cluster fixed point, geometric fixed point, and second percolation fixed point, respectively. 
    In subfigures (a2)-(c2), we zoom in on the regions of the corresponding fixed point.}

    \label{fig:thersholds}
\end{figure*}

In this section, we present numerical results for the scaling behavior along the critical line $K=K_c$, including estimates of the percolation thresholds $p_{c1}$, $p_{c2}$, and their associated critical exponents. Monte Carlo data are analyzed using least-squares fitting based on the finite-size scaling ansatz. We exclude small system sizes by imposing a lower cutoff $L \geq L_{\text{m}}$ to minimize the influence of corrections to scaling.
We systematically examine the effect on the residuals $\chi^2$ by increasing $L_{\text{m}}$. Generally, the preferred fit corresponds to the smallest $L_{\text{m}}$ that yields a reasonable goodness-of-fit and for which further increases in $L_{\text{m}}$ do not cause the $\chi^2$ value to decrease by significantly more than one unit per degree of freedom (DF). In practice, ``reasonable" is defined as having $\chi^2/\text{DF} \approx 1$. 
Systematic uncertainties are quantified by comparing results across alternative fitting ansatz, ensuring robustness against omitted scaling corrections.

\par

To illustrate the phase diagram in Fig.~\ref{fig:phase_diagram}, we consider a line of occupation probabilities $p$ ranging from 0 to 0.8, with system sizes $L = 32, 64, 128$. We plot the critical polynomials $P_{b}$, $P_{bh}$, and $P_{b2}$ versus $p$, as shown in Figs.~\ref{fig:thersholds}(a1)-(c1).

\begin{figure}[ht]
    \centering
    \includegraphics[width=1\linewidth]{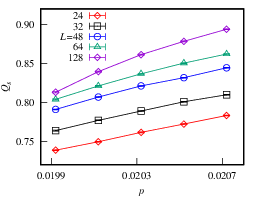}
    \caption{Plots of Binder ratio $Q_{s}$ versus probabilities $p$ near $p_{c1}$ for various system sizes $L$ exhibit stronger finite-size corrections compared to the critical polynomial $P_b$ shown in Fig.~\ref{fig:thersholds}(b2). }
    \label{fig:Qst}
\end{figure}

In Fig.~\ref{fig:thersholds}(a1), the value of $P_{b}$ increases from $-2$ to $1$ as $p$ increases, with a notable intersection near $0.02$, indicated by a blue arrow. 
We zoom in the critical region in Fig.~\ref{fig:thersholds}(a2) and observe the intersection at $ p \approx 0.0203$ and  $P_{b} \approx 0$.
We note that as $L$ increases, the slope of the lines increases, suggesting an unstable fixed point. 
For comparison, we also plot $Q_{s}$ versus $p$ in Fig.~\ref{fig:Qst} with the same system sizes $L$ and range of $p$. 
The intersection of data from various system sizes is not observed in this region.
These results demonstrate that $P_b$ suffers from dramatically less finite-size correction and is more reliable than $Q_{s}$ to extract critical properties. As $p$ increases further, the value of $P_b$ quickly approaches some constant that is very close to the saturated value $P_b=1$. This indicates the definite existence of a large cluster that wraps around in both directions.

We construct different critical polynomials by experimentally testing the combination of various wrapping probabilities. The choice of critical polynomials, $P_{bh}$ and $P_{b2}$, is primarily based on their less finite-size corrections and their approximate values approaching zero at $p_g$ and $p_{c2}$, respectively. This suggests that the design of these critical polynomials effectively captures the intrinsic properties of the fixed points.
In Fig.\ref{fig:thersholds}(b1), the variation of $P_{bh}$ is illustrated as it changes from $-1$ to $1$ with increasing $p$, demonstrating the transition of both majority and minority clusters from not wrapping to wrapping. Around $p \approx 0.036$, a stable fixed point emerges near $P_{bh} \sim 0$, with a detailed view of $P_{bh}$ in this vicinity shown in Fig.\ref{fig:thersholds}(b2).
Figures \ref{fig:thersholds}(c1,c2) depict how $P_{b2}$ evolves with varying $p$ and provide a detailed view of its behavior around the second percolation fixed point. Notably, the existence of two percolation transitions, $p_{c1}$ and $p_{c2}$, is evidenced by the approximate intersections in the plot of $P_{bh}$. We will then discuss the characteristics of each fixed point in detail.

\subsection{First percolation transition}

\begin{table*}
    \caption{Fitting results of $P_{b}$ for bond percolation on the critical Ising model with $z_p=80$ at the random cluster fixed point. The values without error bars mean this parameter is fixed. }
    \begin{ruledtabular}
    \begin{tabular}[t]{l l l l l l l l l l l l l l}
    
        Obs. & $L_{\text{m}}$ & $\chi^{2}/DF$ & $y_{p1}$ & $p_{c1}$ & $\scrO_c$ & $q_{1}$ & $q_{2}$ & $q_{3}$  & $q_{4}$ & $b_{1}$ & $y_{1}$ & $c_{1}$ &  $n_{2}$ \\
    \hline
     $P_{b}$ & 48 & 29.8/25 & 0.543(2) & 0.020~325~8(8) & ~0.000~2(8) &  1.28(1) & -0.75(2)  & -0.096(6) & 0.35(3) & -7(1) & -2 & 48(11) & 0.8(2) \\
     & 64 & 24.4/20 & 0.543(2) & 0.020~325~5(9) & -0.000~3(10) &  1.29(1) & -0.74(3)  & -0.098(7) 
     &0.34(4) &-4(3) & -2 & 42(20) & 0.7(3) \\
       & 48 & 30.2/26 & 13/24     & 0.020~325~6(3) & ~0
       &   1.290(3) & -0.76(2) & -0.08(3) & 0.36(3)
       & -7.0(9)  & -2 & 44(8) & 0.9(2) \\
       & 64 & 24.7/22 & 13/24     & 0.020~325~8(3) & ~0
       &   1.29(1) & -0.75(2) & -0.097(7) &0.34(3) & -5(2) & -2 & 35(11) & 0.7(2) \\
    
    \end{tabular}
    \end{ruledtabular}
    \label{tab:Pbtrc}
\end{table*}

\begin{table}[t]
    \caption{Fitting results of mean size of largest cluster $C_1$  at the random cluster point with $p_{c1} = 0.020\,325\,7$.  }
    \begin{ruledtabular}
    \begin{tabular}[t]{l l l l l l l l }
     Obs. & $L_{\text{m}}$ & $\chi^{2}/DF$ & $y_{h1}$ & $a$ & $b_{1}$ & $b_{2}$ & $y_{1}$  \\
    \hline
    $C_1$ & 24 & 10.7/7 & 1.874~9(5) & 0.595(2) &  0.05(4) & 0 & -0.9(4) \\
    & 32 & 5.7/6 & 1.874~9(5) & 0.594(2) &  0.06(4) & 0 & -0.9(3)  \\
    & 24 & 7.4/7  & 1.875~2(3) & 0.593(1) &  -0.02(4) & 0.04(2)& -1  \\
    & 32 & 5.6/6  & 1.874~9(4) & 0.594(2) &  0.05(7) & 0.01(3)& -1  \\
    \end{tabular}
    \end{ruledtabular}
    \label{tab:C1rc}
\end{table}
    
We utilize the quantity $ P_b $ to determine the random-cluster fixed point $ p_{c1} $ and its associated exponent $ y_{p1} $ for system sizes up to $ L = 512 $, as shown in Fig.~\ref{fig:thersholds}(a2).
We perform the least-square fitting to $P_b(p,L)$ near $p=0.020\,3$ with the following ansatz
\begin{equation}
    \label{Eq:fits}
\begin{aligned}
    \mathcal{O}(t,L)&=\mathcal{O}_c+\sum_{k=1}^m q_k t^k L^{k y_p} + b_1 L^{y_1}\\
    &+c_1t L^{y_1+y_p}+n_2t^2L^{y_p}.
\end{aligned}
\end{equation}
Here, $\scrO_c$ is a constant, $m$ is the highest order retained in the fitting ansatz, and the variable $t$ is defined as $t = (p - p_{c1})/p_{c1}$.
The exponent $y_p$ corresponds to the RG exponent along the $p$ direction, denoted as the $y_{p1}$,  and $y_1 <0 $ is the correction exponent. 
The term $c_1tL^{y_1 + y_p}$ accounts for the crossing effect between the finite-size corrections and the scaling variable $tL^{y_p}$, and the $n_2$ term originates from the nonlinear component of the RG invariant function.

\begin{figure}[ht]
    \centering
    \includegraphics[width=1\linewidth]{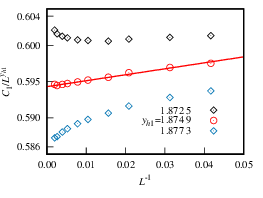}
    \caption{Plots of the rescaled largest cluster $C_1/L^{y_{h1}}$ versus  $L^{-1}$. The upper and lower bending tendencies represent scaling with $y_{h1}$ values that deviate from the estimated central value by three times the quoted error.}
    \label{fig:random}
\end{figure}

The fitting analysis of $ P_b $ is conducted using Eq.\eqref{Eq:fits}.
Our preliminary fits reveal that simultaneously fitting both $y_{p1}$ and $y_1$ produces unstable outcomes, though $ y_1 $ consistently approaches -2. 
Additionally, we observe that the coefficients $q_k$ become negligible for $k>4$. 
Consequently, we fix $ y_1 = -2 $ and retain $ m = 4 $. We then allow all other parameters to be fitted freely and obtain $ p_{c1} = 0.020\,325\,6(10) $ and $ y_{p1} = 0.543(2) $, which is consistent with the red-bond exponent $ y_r = 13/24 $ of the FK Ising model. Furthermore, the constant $ \mathcal{O}_c $ is consistent with 0, aligning with the observation $ \lim_{t \to 0 } P_b(t, L) = 0 $ for the FK Ising model. These findings suggest that the percolation transition at $p_{c1}$ belongs to the critical FK Ising universality class. For further refinement, we perform an additional fit by fixing $ y_{p1} = 13/24 $ and $ \mathcal{O}_c = 0 $. These fits stabilize at $ L_{\text{m}} = 48 $ and yield a more precise estimate of $ p_{c1} = 0.020\,325\,7(5) $.
Comparing with various fitting ansatzs, we obtain the final estimates as  $ p_{c1} = 0.020\,325\,7(5) $ and $ y_{p1} = 0.543(3) $, as presented in Table~\ref{tab:Pbtrc}.  

We further determine the fractal dimension of the largest cluster, denoted as $ y_{h1} $. Keeping the bond probability at $ p = p_{c1} = 0.020\,325\,7 $, we measure the largest cluster size $ C_1 $, which is expected to scale as $ C_1 \sim L^{y_{h1}} $. We then try to extract the critical exponent $y_{h1}$ through the ansatz 
\begin{equation}
\label{Eq:fitC1rc}
  \scrO  = L^{y_{h}}( a+b_{1}L^{y_{1}} + b_2 L^{y_2} )+c_0.
\end{equation}
Initially, we leave all parameters free, which gives unstable results. We then fix $ y_2 = 2y_1 $ and $ c_0 = 0 $, which yields $y_{h1} = 1.847\,9(5)$ and $ y_1 = -0.9(4) $ for $L_{\text{m}}=24$. 
Based on it, we fix $ y_1 = -1 $ and  obtain the estimate  $ y_{h1} = 1.874\,9(8) $. This value shows excellent agreement with the theoretical prediction $ y_{h1} = 15/8 $ for the FK Ising model. 
The fitting results are summarized in Table~\ref{tab:C1rc}. In Fig.~\ref{fig:random}, we plot the rescaled largest cluster $ C_1/L^{y_{h1}} $ versus $ L^{-1} $, with $ y_{h1} $ taking the central value of the estimate as well as the central value plus or minus three error bars.
The approximate linearity of the red line and the upward and downward bending curves reflect the reliability of
the final estimate $ y_{h1} = 1.874\,9 $ and the quoted error margin of $ 0.000\,8 $.

In short, we find that the asymptotic value of the critical polynomial $ P_b(p \to p_c, L \to \infty) $ and the critical exponents $ y_{h1} $ and $ y_{p1} $ are consistent with those of the critical FK Ising model, indicating that the universality class does not change for $z_p = 80$.

\subsection{Geometric fixed point}

\begin{figure}[ht]
    \centering
    \includegraphics[width=1\linewidth]{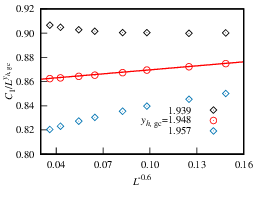}
    \caption{Plots of the rescaled largest cluster $C_1/L^{y_h}$ versus $L^{-0.6}$. The upper and lower bending trends correspond to scaling with \( y_{h} \) values that deviate from the central estimate by three times the quoted error.}
    \label{fig:Geometric}
\end{figure}

We then consider the geometric fixed point.
The data for $ P_{bh} $ are presented in Fig.~\ref{fig:thersholds}, which gives a roughly common intersection near $p_{g}=0.035$ with $P_{bh} \approx 0$. Notably, the slopes of the $ P_{bh} $ data as a function of $ p $ decrease and approach zero as the system size $ L $ increases, suggesting that the bond-placing operator becomes irrelevant at the quasi-stable fixed point $ p_g $. It is somewhat surprising that the intersection for various $ L $ values is so precise, given that various irrelevant scaling fields should simultaneously exist. The exact value $ P_{bh} = 0 $ can be derived from the self-matching property when the percolation game is played on a triangulated lattice with coordination number $ z_p = 6 $. 

Using a similar procedure to the analysis of the first percolation transition, we estimate $ p_g = 0.035\,3(6) $ and $ y_{p,\mathrm{gc}} = -0.624(1) $ through finite-size scaling analyses of $ P_{bh} $. These results are consistent with the predicted geometric fixed-point exponent of $-5/8$. At $ p = p_g $, we determine the fractal dimension of the largest cluster to be $ y_{h,\mathrm{gc}} = 1.948(3) $, which agrees with the spin-cluster fractal dimension $ 187/96 $~\cite{stella1989scaling}. In Fig.~\ref{fig:Geometric}, we take a similar procedure to present the reliability of the estimate of the central value and quoted error of $y_{\rm{h, gc}}$. 
Finally, we emphasize that the critical value $ P_{bh} = 0 $ and critical exponents, such as $ y_{h,\mathrm{gc}} = 187/96 $, remain valid across the entire range $ p_{c1} < p < p_{c2} $.

\subsection{Second percolation transition }

\begin{table*}[ht]
    \caption{ Finite-size scaling analysis of the critical polynomial $P_{b2}$ for equivalent-neighbor percolation clusters on critical 2D Ising spin configurations with $z_p=80$ at the second percolation fixed point. Results are obtained through least-squares fitting to the ansatz in Eq.~\eqref{Eq:fits}. The table presents the estimated critical parameters and their uncertainties as a function of the minimum system size $L_{\text{m}}$ included in the fit. }
    \tabcolsep=0.26 cm
    \begin{tabular}{lllllllllll}
    \hline
    \hline
    Obs. & $L_{\text{m}}$ & $\chi^2$/DF & $y_{p2}$ & $p_{c2}$  & $\mathcal{O}_c$ & $q_1$&  $q_2$& $b_1$ & $y_1$   \\
    \hline
    $P_{b2}$  
    & 16 & 54.0/51 & 0.553(6) & 0.345\ 81(1) & 0.001\ 6(1) & 0.155(4) & 0.050(17)  & -3.1(4) & -1.99(4)\\
    & 32 & 49.6/46 & 0.551(8) & 0.345\ 81(2) & 0.001\ 7(2) & 0.158(6) & 0.052(18)  & -2(1) & -1.9(2)\\
    & 16 & 54.0/52 & 0.553(6) & 0.345\ 80(1) & 0.001\ 68(8) & 0.155(4) & 0.050(17)  & -3.15(2) & -2\\
    & 32 & 49.7/47 & 0.552(7) & 0.345\ 81(1) & 0.001\ 7(1) & 0.156(5) & 0.053(18)  & -3.2(1) & -2\\
    & 16 & 57.4/53 & 13/24 & 0.345\ 82(1) & 0.001\ 7(1) & 0.163~9(6) & 0.044(18)  & -3.1(4) & -2.00(4)\\
    & 32 & 51.0/47 & 13/24 & 0.345\ 82(2) & 0.001\ 9(2) & 0.164~2(7) & 0.053(21)  & -1.9(12) & -1.9(1)\\
    \hline
    \hline
    \end{tabular}

    \label{table:Pb2}
\end{table*}

Finally, we investigate the second percolation threshold, denoted as $ p_{c2} $. As depicted in Fig.~\ref{fig:thersholds}(c2), we utilize the critical polynomial $ P_{b2} $ to determine $ p_{c2} $ and its associated critical exponent $ y_{p2} $. Our methodology involves applying least-squares fitting to the finite-size scaling ansatz specified in Eq.~\eqref{Eq:fits}. Initially, we perform fits with all parameters free.
This preliminary analysis indicates that the coefficients $ c_1 $ and $ n_2 $ are effectively zero. Further, we refine our procedure by setting $ c_1 =  n_2 = 0 $, which gives the estimate $ p_{c2} = 0.345\,81(1) $, $ y_{p2} = 0.553(11) $, and a correction exponent $ y_1 = -1.99(4) $. Meanwhile, the critical value of $ P_{b2} $ is very close to zero but not precisely zero.  These results are summarized in Table~\ref{table:Pb2}. \par 

\begin{table}[ht]
    \caption{Finite-size scaling analysis of the mean size of the largest clusters $C_1$ at second percolation fixed point $p_{c2}$ for equivalent-neighbor percolation with $z_p=80$. Results are obtained through least-squares fitting to the ansatz in Eq.~\eqref{Eq:fitC1rc}.}
    \centering
    \tabcolsep=0.07 cm
    \begin{tabular}{llllllll}
    \hline
    \hline
    Obs. & $L_{\text{m}}$ & $\chi^2$/DF & $y_{h2}$ & $a$&  $b_1$ & $c$ & $y_1$   \\
    \hline
    $C_1$  
    & 32 & 1.3/5 & 1.958~0(1) & 0.945~8(7) & 0.47(1) & -3.7(2) & -1 \\
    & 48 & 1.2/4 & 1.958~0(2) & 0.946(1) & 0.46(5) & -3(1) & -1 \\
    \hline
    \hline
    \end{tabular}
    \label{tab:C1}
\end{table}

We then set $ y_1 = -2 $, which provides estimates of $ y_{p2} = 0.552(9) $ and $ p_{c2} = 0.345\,82(2) $. Notably, the value of $ y_{p2} $ is remarkably close to $ y_{p1} = 13/24 $.
We then fix $y_{p2}=13/24$, which yields consistent estimates for other parameters. 
To ensure the stability and reliability of our estimates, we increase the minimum system size $ L_{\text{m}} $ included in our fits. By comparing various results, we derive final estimates of $ y_{p2} = 0.552(9) $ and $ p_{c2} = 0.345\,82(2) $. The remarkable alignment between $ y_{p2} $ and the theoretically predicted value for the random-cluster fixed point suggests a potential underlying mechanism in which there probably exists some kind of duality between $p_{c1}$ and $p_{c2}$, which needs future theoretical exploration.

\begin{figure}[ht]
    \centering
    \includegraphics[width=1\linewidth]{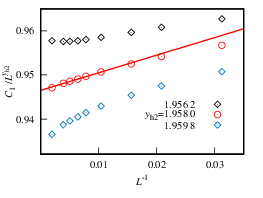}
    \caption{Estimation of the magnetic exponent $y_{h2}$ at the second percolation fixed point. Scaled largest cluster size $C_{\text{1}}/L^{y_{h2}}$ against $L^{-1}$. The upward and downward bending tendencies represent the reliability of the final quoted central values.}
    \label{fig:C1}
\end{figure}

After determining the exponent $y_{p2}$ from the critical polynomial $P_{b2}$, we further extracted the magnetic exponent $y_{h2}$ through finite-size scaling analysis of $C_1$ at $p = p_{c2}$ via  the ansatz Eq.~\eqref{Eq:fitC1rc}. 
Initially, all parameters vary freely, and we obtain unstable results. Subsequently, we adopted a more constrained strategy by fixing $y_1 = -1$ while keeping the other parameters free. This strategy produced highly consistent results for $L_{\text{m}} \geq 32$, yielding $y_{h2} = 1.958~0(2)$. To ensure robustness, we compare fits across various values of $y_{h2}$ and $L_{\text{m}}$, ultimately obtaining a refined estimate of $y_{h2} = 1.958~0(6)$ by accounting for systematic errors. Details of the fitting procedure are summarized in Table~\ref{tab:C1}. Figure~\ref{fig:C1} presents a plot of $C_1/L^{y_{h2}}$ against $L^{-1}$, displaying the central estimate of $y_{h2}$ alongside values at plus and minus three standard errors. The near-linear progression of the central red line, together with the curvature of the error bound lines, corroborates our estimate of $y_{h2} = 1.9580$ with a quoted error of $0.000~6$.

\begin{figure}[ht]
    \centering
    \includegraphics[width=1\linewidth]{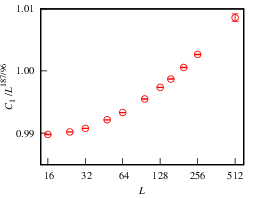}
    \caption{Log-log plot of $C_1/L^{187/96}$ versus system size $L$ at the second percolation threshold $p_{c2}$ for equivalent-neighbor percolation with $z_p=80$. The upward trend of the data points with increasing $L$ demonstrates that the magnetic exponent $y_{h2}$ is larger than 187/96. }
    \label{fig:C1fit}
\end{figure}

\begin{figure}[ht]
    \centering
    \includegraphics[width=1\linewidth]{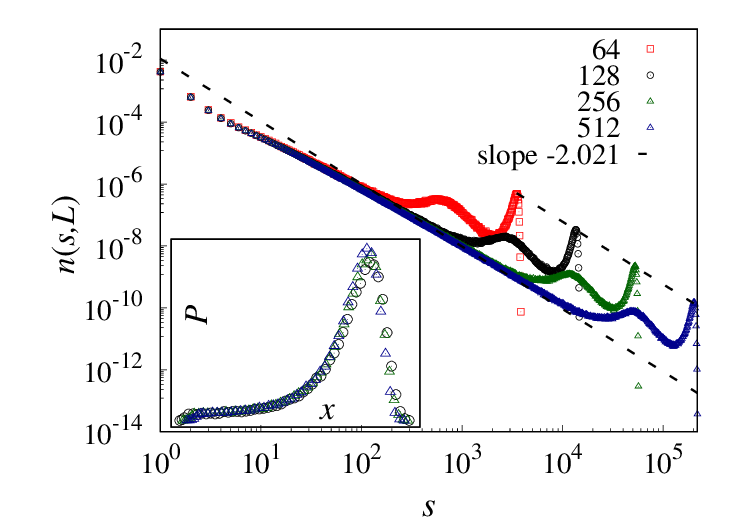}
    \caption{Log-log plot of the cluster-size distribution $n(s,L)$ versus $s$ for $L=64,128,256$ and $512$ at the second-percolation point. The black dashed line has a slope of $-\tau=-2.021$, indicating the validity of the hyperscaling relation. In the inset, we plot the probability of the rescaled largest cluster as $x = C_1/L^{y_{h2}}$. The good data collapse demonstrates the validity of the estimate $y_{h2} = 1.958\,0(6)$. }
    \label{fig:ns}
\end{figure}

We note that $y_{p2} = 0.552(9)$ is consistent with the red-bond exponent $y_r = 13/24$, as well as with $y_{p1}$. This raises the question of whether the corresponding RG exponents at the s.p. fixed point are related to these known exponents.
One conjecture is that $y_{h2} = 1.958~0(6)$ may align with the fractal dimension at the geometric fixed point, specifically $187/96 = 1.947~92...$. Figure~\ref{fig:C1fit} presents a log-log plot of $C_1/L^{187/96}$ against the system size $L$. As $L$ increases, $C_1/L^{187/96}$ exhibits a marked upward trend and does not converge to a constant even up to $L = 512$, suggesting that $y_{h2}$ is larger than $187/96$. The exact value of $y_{h2}$ requires further investigation.  

Finally, we consider the cluster number density $n(s,L)$ at $p_{c2}$. At criticality, it follows the relation:
    \begin{equation}
n(s,L)=s^{-\tau}\tilde{n}(s/L^{d_{\textsc{f}}})\quad [\tilde{n}(x\rightarrow 0)=1],
\label{Eq:ns}
\end{equation}
where $\tilde{n}(x)$ is a universal scaling function, and the Fisher exponent $\tau$ satisfies the hyperscaling relation $\tau = 1 + d/d_{\textsc{f}}$ with the fractal dimension of the largest cluster $d_{\textsc{f}}$. Figure~\ref{fig:ns} shows a log-log plot of $n(s,L)$ against $s$, initially revealing a power-law scaling behavior with a slope consistent with $-2.021$, corroborating the hyperscaling relation $\tau = 1 + 2/1.958 = 2.021$. Subsequently, the decay rate of $n(s,L)$ slows, displaying a non-monotonic pattern due to the probability distribution of the largest cluster. In the inset of Fig.~\ref{fig:ns}, we plot the probability distribution of the largest cluster $P(x)$, where $x = C_1/L^{y_{h2}}$.
The data from various system sizes collapse well, affirming the validity of the estimate for $y_{h2}$.

\section{Discussion}
\label{Sec:discussion}
We perform Monte Carlo simulations on the square-lattice Ising model at criticality and 
construct extended-neighbor percolation clusters with a coordination number $z_p = 80$. Through various critical polynomials, we obtain the critical exponents $y_{p1}=0.543(3)$ and $y_{h1}=1.874~9(8)$ at the first percolation point, and $y_{p, \rm{gc}}=-0.624(1)$ and $y_{h, \rm{gc}}=1.948(3)$ at the geometric cluster fixed point. These results are consistent with the theoretical predictions from the Coulomb gas theory.
We further observe a second percolation threshold with the corresponding renormalization group exponents estimated as $y_{p2} = 0.552(9)$ and  $y_{h2} = 1.958\,0(6)$, where the value of $y_{p2}$ is numerically consistent with the red-bond exponent $y_r=13/24$ of $y_{p1}$, and the value of $y_{h2}$ does not correspond to any  known exponents for two-dimensional Ising model.

Since the percolation clusters are constructed over an extended range with $z_p=80$, the exact FK transformation from the spin to the random-cluster representation of the Ising model no longer exists. As a consequence, the first percolation threshold $p_{c1} \approx 0.020$ for the extended percolation is much smaller than $p_{ci} = 2-\sqrt{2}\approx 0.58$ for the critical FK-Ising model on the square lattice with coordination number $z_i=4$. Despite of this dramatic difference, the critical behaviors of percolation clusters at $p_{c1}$ agree well with those for the critical FK random clusters, and the same correspondence holds between the stable critical phase in range $p_{c1} <p <p_{c2}$ and the spin domains (droplets) of the Ising model. This provides a beautiful illustration of the robustness of universality, a core concept in the modern theory of critical phenomena.

Based on the universality argument, we expect that the overall structure of the phase diagram in 
Fig.~\ref{fig:phase_diagram} holds generally true in two dimensions. Namely, it depends neither on the interaction range 
for the Ising model nor on the bond-occupation range for percolation, as long as they are within 
the short-range regime. This is indeed supported by our additional simulations for the extended 
Ising model—e.g., with coordination number $z_i=24$, and for percolation ranges besides $z_p=80$. 
The critical behaviors at the two percolation thresholds as well as for the stable critical phase in
between remain the same as in the current case ($z_i$= 4 and $z_p$=80), and, thus, they are not shown. 
In three and higher dimensions, however, the percolation properties of the Ising model are 
dramatically different from those in two dimensions. For instance, at the critical point, the Ising
domains of the up and down spins are giant instead of being fractal. By modifying the flexible
percolation rule—e.g., placing bonds only within a single layer of Ising configurations--one might 
hope to obtain rich and insightful geometric properties for the high-dimensional Ising model. It
would also be interesting to study percolation in two- and multi-layered Ising models that are still 
in the two-dimensional universality, which could reveal interesting crossover phenomena from 
two to three dimensions.

\section*{Acknowledgments}  This work was supported by the National Natural Science Foundation of China under Grant No. 12275263
(Y.D.), the Innovation Program for Quantum Science and Technology under Grant
No. 2021ZD0301900, and the Natural Science Foundation
of Fujian Province of China under Grant No. 2023J02032.

\bibliography{references.bib}

\end{document}